\documentclass[english]{article}
\usepackage[T1]{fontenc}
\usepackage[latin9]{inputenc}
\usepackage{graphicx}

\makeatletter
\makeatletter
\makeatletter
\usepackage{multicol}

\makeatother

\makeatother

\makeatother

\usepackage{babel}

\begin{document}
\begin{center}
\textbf{\Large Turbulent viscosity variability in self-propelled body
wake model.}{\Large{} }
\par\end{center}{\Large \par}

\begin{center}
\textbf{$^{1}$Katya Dubrovin, $^{1}$Michael Gedalin, $^{1}$Ephim
Golbraikh}\\
\textbf{ and $^{2}$Alex Soloviev}
\par\end{center}

\begin{center}
$^{1}$Physics Department,Ben-Gurion University of the Negev, Beer-Sheva,
Israel\\
 $^{2}$Oceanographic Center, NOVA SouthEastern University, USA
\par\end{center}

\begin{abstract}
We study the influence of turbulent viscosity variability on the properties
of self-propelled body wake model. In addition to the already known
integrals of motion obtained with constant turbulent viscosity, we
obtain new ones. The presence of new integrals of motion leads, in
particular, to changes in the behavior of the width and profile of
the wake leading to its conservation.
\end{abstract}
Study the behavior of long-lived turbulent wakes behind moving bodies
is not only important from a scientific \cite{key-2,key-3,key-13}
but also practical point of view \cite{key-4,key-7,key-10}. On the
one hand the evolution of such wakes can affect the operation of airports,
and on the other - mixing processes in the surface layer of the sea
or ocean. The turbulent wakes behind ships have attracted special
interest, largely due to the need to interpret the radar observations
of ship wakes \cite{key-8,key-10,key-11,key-12}.

One of the most important models of the turbulent wakes behavior,
from a practical point of view, is the model of self-propelled body
wake \cite{key-2,key-5,key-6,key-7,key9}, and the main characteristic
of this wake is turbulent viscosity,$\nu_{t}$ . Analytical approaches
so far assumed a constant viscosity $\nu_{t}=const$ , which leads
to the wake width at the distance x equal to $W(x)\sim x^{\alpha}$(where
$\alpha=1/5$ for an axisymmetric wake and for $\alpha=1/4$ a plane
one). However, in real wakes, turbulent viscosity may be a function
of the coordinates. In this paper we study the consequence of this
dependence.

Consider the axisymmetric wake in which the mean flow in planes passing
through the axis of symmetry is identical and weakly depends on the
distance x. Simultaneously, we examine flows whose principal mean
velocity component is directed along the x-axis. This means that radial
component of the mean velocity $\mid U_{r}\mid<<\mid U_{x}\mid$.
If $\mathit{L}$ is a typical spatial scale of inhomogeneity in the
direction of the wake (x -direction) while $\mathit{l}$ is a typical
spatial scale of inhomogeneity in the transverse direction (r), $l<<L$(in
this paper we use the notations in conformity with \cite{key-13}).
Then we denote by the mean flow velocity in x direction, and by $U_{s}$
the maximum value of $\mid U_{0}-U_{x}\mid$ ($U_{s}<<U_{0}$ far
from the body, ) of the cross-wake variation of the mean velocity
component in direction. Following \cite{key-13}, an equation for
the momentum transfer in x direction can be written as: \begin{equation}
U_{0}\frac{\partial\tilde{U}}{\partial x}+\frac{1}{r}\frac{\partial rq_{xr}}{\partial r}=0\label{eq:1}\end{equation}
 where $\tilde{U}=U_{x}-U_{0}$ , $q_{xr}=<u_{x}u_{r}>$, $u_{i}$
is a turbulent component of the velocity field, and <...> denotes
averaging over an ensemble. Taking into account self-preservation
hypothesis,: the velocity defect and the Reynolds stress become invariant
with respect to x, and they are expressed in terms of the local length
and velocity scales $\mathit{l(x)}$ and $\mathit{U_{s}(x)}$ . That
is, $\mathit{\frac{\tilde{U}}{U_{s}}=f(r/l)}$ and $\mathit{\frac{q_{xr}}{U_{s}^{2}}=-g(r/l)}$.
We substitute these expressions into the equation of motion (\ref{eq:1}):

\begin{equation}
-\frac{U_{0}l}{U_{s}^{2}}\frac{dU_{s}}{dx}f+\frac{U_{0}}{U_{s}}\frac{dl}{dx}\xi f^{'}=g^{'}+\frac{g}{\xi}\label{eq:2}\end{equation}
 where $\xi=r/l(x)$. To satisfy this equation for all x we obtain
$l=Ax^{\alpha},U_{s}=Bx^{\alpha-1}$, where $A$ and $B$ are constants.
Substitution into (\ref{eq:2}) gives

\begin{equation}
s(\alpha-1)f-s\alpha\xi f^{\prime}=\frac{1}{\xi}\frac{d}{d\xi}(\xi g)\label{eq:3}\end{equation}
 where $s=(U_{0}/B)$. The functions $f(\xi)$ and $g(\xi)$ are still
to be found.

Since the turbulent viscosity is defined by the relation

\begin{equation}
-q_{xr}=\nu_{t}\frac{\partial\tilde{U}}{\partial r},\label{eq:visc}\end{equation}
 the equation of motion acquires the following form:

\begin{equation}
U_{0}\frac{\partial\tilde{U}}{\partial x}=\frac{1}{r}\frac{\partial}{\partial r}\left(r\nu_{t}\frac{\partial\tilde{U}}{\partial r}\right)\label{eq:4}\end{equation}
 where $\nu_{t}=Bx^{2\alpha-1}(g/f^{\prime})$, and prime denotes
$\xi$ derivative.

The integral $I_{0}=\int\limits _{0}^{\infty}\tilde{U}rdr$ is proportional
to the total momentum across the wake and is conserved. Indeed, $\frac{d}{dx}\int\limits _{0}^{\infty}U_{0}\tilde{U}rdr=-\int\limits _{0}^{\infty}\frac{\partial}{\partial r}\left(rq_{xr}\right)dr=0$.
The conservation of momentum, together with the assumption that $\tilde{U}\propto xf(\xi)$,
immediately gives $I_{0}\propto x^{3\alpha-1}\int\nolimits _{0}^{\infty}f(\xi)\xi d\xi$,
and therefore we obtain a well-known value $\alpha=1/3$.

However, it fails for self-propelled body wake, where the total momentum
vanishes ($I_{0}=0$) . In this case, multiplying by $r^{m+1}$ and
assuming the convergence, one has $\frac{d}{dx}I_{m}=-\int\nolimits _{0}^{\infty}r^{m}\frac{\partial}{\partial r}(rq_{xr})dr$
, where $I_{m}=\int\limits _{0}^{\infty}U_{0}\tilde{U}r^{m+1}dr$.
Using turbulent viscosity, we obtain:

\[
\frac{d}{dx}I_{m}=\int\limits _{0}^{\infty}r^{m}\frac{\partial}{\partial r}\left(r\nu_{t}\frac{\partial\tilde{U}}{\partial r}\right)dr\]
 Integrating in parts twice and taking into account that $\tilde{U}\rightarrow0$
and $\frac{\partial\tilde{U}}{\partial r}\rightarrow0$ as $r\rightarrow\infty$
:\[
\frac{d}{dx}I_{m}=m\int\limits _{0}^{\infty}\tilde{U}\frac{\partial}{\partial r}\left(r^{m}\nu_{t}\right)dr\]
 If $\frac{\partial}{\partial r}(r^{m}\nu_{t})\propto r$ then $\int\limits _{0}^{\infty}\tilde{U}\frac{\partial}{\partial r}\left(r^{m}\nu_{t}\right)dr\propto I_{0}=0$and
$I_{m}$ is conserved.

In this case, $\nu_{t}=K(x)\xi^{2-m}$, where $\mathit{K(x)}$ is
some function on x. Conservation of $I_{m}$ , together with self-preservation
hypothesis, gives \begin{equation}
\alpha=\frac{1}{m+3}\label{eq:alpha}\end{equation}
 It should be noted that similar results can be obtained for the plane
flow. In this case we obtain $\alpha=\frac{1}{m+2}$ .

Thus, with turbulent viscosity definition and conserved $I_{m}$,
we obtain

$K(x)\xi^{2-m}=Bx^{2\alpha-1}g/f^{\prime}$, and therefore $K(x)=Bx^{2\alpha-1}$
and $g=\xi^{2-m}f^{\prime}$ .The latter relation can be now substituted
into (\ref{eq:3}). Eventually, we arrive at the following equation:

\begin{equation}
f^{\prime\prime}+\left[\frac{3-m}{\xi}+\frac{s}{m+3}\xi^{m-1}\right]f^{\prime}+\frac{s(m+2)}{m+3}\xi^{m-2}f=0\label{eq:5}\end{equation}

The solution of this equation for certain values of $\mathit{m}$
(and $s=1$) is shown in Fig. 1.

\begin{figure}
\includegraphics[width=4in]{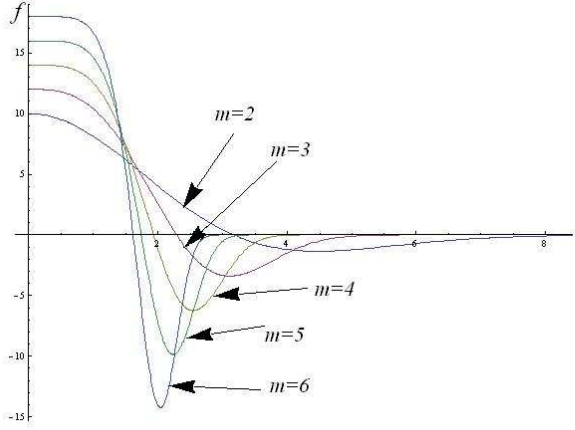}\\

\caption{Wake profiles for certain parameters $\mathit{m}$. $\mathit{m=2}$
corresponds to $\mathit{\nu_{t}=const}$.}

\end{figure}

It is worth mentioning that although the turbulence viscosity coefficient,
formally defined by (\ref{eq:visc}), diverges as $\xi\rightarrow0$
, physical quantities remain well- defined. Really, $\tilde{U}(0)$
is constant and nonzero, near $\xi=0$ :  $f^{'}\propto\xi^{m-1}exp(-a\xi^{m})$ (where \emph{a} is a constant)
, and $q_{xr}\propto g\propto\xi^{2-m}f^{\prime}\propto\xi$ .

As we can see in Fig. 1, the growth of $\mathit{m}$ (more rapid decrease
of the turbulent viscosity with $\xi$) leads to an effective reduction
of the wake width. At the same time the zone characterized by $\tilde{U}/U_{s}=const$
increases.

Thus, in the present work we have suggested to get rid of the assumption
of the uniform turbulent viscosity for a self-propelled body wake.
Instead, we consider a power-law dependence $\nu_{T}\propto r^{2-m}$,
which allows us to construct a new integral of motion, $I_{m}=\int\nolimits _{0}^{\infty}r^{m}\tilde{U}rdr$
instead of the vanishing $I_{0}=\int\nolimits _{0}^{\infty}\tilde{U}rdr$.
Each of these integrals leads to the formation of a corresponding
flow profile. Simultaneously, the increase in the parameter $\mathit{m}$
leads to an increase in the flow \char`\"{}core\char`\"{} and slower
growth of its width.

\end{document}